# NUMERICAL MODELING OF ROCKMASS BEHAVIOUR DUE TO BLASTING OPERATIONS IN UNDERGROUND MINES

**Witold Pytel** [1], **Piotr Mertuszka** [2], **Krzysztof Fuławka** [2], **Marcin Szumny** [2], **Lech Stolecki** [2]

[1] Wrocław University of Science and Technology, **Poland**
[2] KGHM CUPRUM Ltd. Research and Development Centre, **Poland**

**ABSTRACT**

*Antropogenic seismicity is one of the most important geomechanical hazard encountered in Polish deep copper mines. Obviously, it may have a negative impact on the people's safety and the production process. In such circumstances, a number of both active and passive prevention measures have been implemented in the mines, at least partially. One of the most effective, applied for many years, active means used for seismic hazard reduction, is multi-faces blasting works which by assumption should result in release of strain energy accumulated in the rock mass by inducing seismic events in the areas of high strain concentrations. However, at the stage of blasting works design process, a most of attention is paid rather on the vibration amplitude increasing than on the effect of its frequencies. In result, one of the basic parameters of induced seismic waves, which is the frequency of vibrations is usually omitted. Nowadays, the frequency of the seismic wave initiated by firing of group of faces is basically out of control due to the low accuracy of detonators utilized today. Taking the above into account, the possibility of increasing the effectiveness of blasting provocation by controlling the frequency of induced seismic waves was analysed as the effect of the frequencies' values assumed to be the dominating ones, around which the computer based 3D FEM solutions of the blasting operation has been simulated.*

1. **INTRODUCTION**

Copper ore extraction in Poland is focused on the deep deposit which is located at south west part of the country and lies at depth up to 1,500 m. The ore body is situated at the contact of three different rock formations, including dolomite, shale and sandstone. This is one level deposit classified as stratoidal formation which declines at angle about 3°. Thickness of the ore body varies from 1.6 m up to 5.0 m. The roof strata consists mainly of dolomite with thickness up to 60 m. In some parts of the ore body, the rock layers are unsettled by faults. The average copper content in extracted ore is about 2%, but the shale has the highest copper mineralisation among other rock layers and can reach up to 30%.

This kind of the deposit structure determine the room-and-pillar extraction system as the most effective mining method. This excavation system is already used for more than 50 years and all the time is developed and adjusted to the current local geomechanical conditions. Changes cover many of mining parameters like pillars geometry, workings height, type of roof support, backfilling etc. The optimization process has to take into considerations many various aspects in which, the most important is to maintain or improve safety level. Nowadays, the room-and-pillar method with controlled roof deflection is commonly used. In some specific mining panel there is hydraulic backfilling used as well. The roof deflection rate must be adjusted to the progress of mining front to avoid dangerous events like i.e. roof fall [1]. Unfortunately, due to specific rocks structure and parameters like thickness and strength, exploitation depth (more than 1,000 m) and other mining conditions, this method is not sufficient enough to avoid seismic events which can cause dangerous phenomena like rockburst. These kind of threats are very dangerous for the crew during presence in the affected area.

In order to reduce the rockburst hazards, many different active and passive prevention methods were implemented to minimize the risk caused by the seismic activity. Passive methods cover mainly work organisation like limitation of the working time and maximum number of the crew in the dangerous area. Unfortunately, this kind of prevention is not fully sufficient and has to be supported by other methods. The most effective preventive method, that is currently used, group blasting works have revealed themselves. The method of determining the effectiveness of distressing blasting is based on analysis of the seismicity recorded by seismic stations. Some seismic events are considered to be "spontaneous", what means, that they are not connected directly to mine production blasting. Other seismic events occur in the time period immediately after production group blasting and therefore they are treated as the

"provoked" seismicity. It is believed that the more of seismic events are "provoked", the more effective group blasting may be considered as.

The basis of the multi-face blasting is firing more than 10 faces simultaneously at the one blasting sequence. Based on the long time on-site observations it was concluded, that the average provocation level depending on area and generally varied between 20-30%, except 2015. It means, that there is still a room for developing such methods. Figure 1 shows the proportion of spontaneous and provoked events in Polish copper mines.

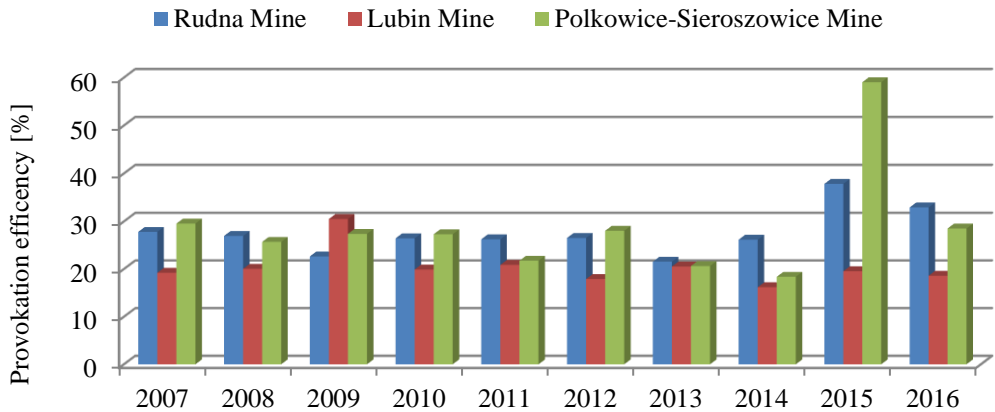

**Fig. 1. Proportion of spontaneous vs. provoked events recorded in Polish copper mines, 2007-2016**

The main controlled factors in the group blasting works is the number of simultaneously fired faces and consequently the amount of detonated explosives. It must be pointed out that blasting works permit achieving some production targets such as required burden, appropriate rock fragmentation and proper level of paraseismic vibrations. It is done by means of using delays detonators with proper drilling and blasting pattern. Currently, the most often used detonators in the underground applications are non-electric detonators, which are precise enough from the exploitation point of view but not enough from the controlling of seismic wave. The accuracy of the most precised non-electric detonators is about ±1%, but in case of typical detonators this value is lower and may reach a ±3-5% from the nominal value [2]. This kind of disadvantage do not reveal electronic initiations systems. Modern electronic detonators are characterised by, among others, two main features, i.e. they are precise and programmable. Level of accuracy reaches ±0.005% and they are fully programmable in the time range from 0 ms up to 30 s in 1 ms increments in the latest generation of electronic detonators. It creates new possibilities in the field of the seismic wave control in wider spectrum, i.e. amplitude, frequency. Many of measurements made in different conditions shown that the time delays and blasting sequence can affected the amplitude and frequency spectrum of inducted seismic waves. Some trials which was done in the underground conditions shown that there are possibilities for amplifying seismic waves by means of proper time sequence made by the electronic detonators [3].

In this article a theoretical numerical model 3D FEM was used to verify the effect of seismic wave frequency on the rock mass behavior. The problem solution was obtained using the 3D FEM numerical model based on the geometry of the selected mining panel in one of the Polish copper mines. Outcomes of these analysis will allow to look into the phenomenon in a broader scope and indicate directions for further field trials where precisely described and selected time sequence could be implemented and checked on site. Therefore seismic waves was analysed from point of view of the effect of the frequencies' values assumed to be the dominating ones, around which the computer based 3D FEM solutions of the blasting operation has been simulated.

## 2. MATERIALS AND METHODS

The measurements of vibration in the rock involve displacement, velocity and acceleration of the particle. All of these quantities are the vectors and normally may be described into three orthogonal components x, y and z. However vibration velocity is the parameter which is most often used in practice for the vibration measurement. This is due to the fact that the velocity for the range of frequencies from 15 Hz up to 1 kHz is today very well suited to measurement capabilities of the commonly utilized geophysical equipment (e.g. seismometers) [4]. Since the problem will be analyzed for scenarios within the mine field inside 3D described rock mass, the vector used for characterizing vibration velocity in each measurement points will be also defined in three dimensions. The particle velocity total value may be then calculated from the following formula:

$$V_{xyz}(t) = \sqrt{V_x^2(t) + V_y^2(t) + V_z^2(t)} \tag{1}$$

On the other hand, the velocity vector may be used for the stress calculation in the rock mass which are produced by the seismic wave. The seismic stress or strain may be roughly estimated on the base of vibration velocity:

$$\varepsilon = \frac{\sigma}{E} \approx \frac{v}{c} \tag{2}$$

where: $\varepsilon$ – strain, $E$ – Young modulus, $v$ – particle velocity, $c$ – stress wave velocity.

Taking into account that some solid rocks may fail in tension at stress of about 3.0 MPa, the corresponding particle velocity may reach 0.2 to 1.0 m/s [5]. In average conditions i.e in the far-field area, seismic waves induced by blasting works do not reach these values of the vibration velocity, however even lower values and proper frequency characteristic may be a cause of small cracks development and thus will decrease the rock strength. From the analysis of several vibrations records triggered by blasting works, carried out within area of Polish copper mines one may conclude, that the range of rock mass dominant frequencies varies from 5 up to 100 Hz [6]. The high frequencies are attenuated much faster than lower so dominant frequencies decrease with the distance from the source of vibrations. For comparison in a case of earthquakes dominant frequencies generally do not exceed a few Hz. It means, that distance and damping factor effectively eliminate the higher frequency band. Taking also into account that the lower frequencies have stronger impact on the rock body in terms of the rock structure damage, it was assumed, that numerical simulations of production blasting focused on the following frequencies: 10, 20 and 50 Hz. Numerical simulations has been first approximation of the impact of the blasting work developing in the direction of the rock structure at distance of about 40÷60 m above the immediate roof strata. The numerical experiment has been conducted in relation to the dominant frequencies of the vibrations induced by blasting works. Results of numerical modeling was described on the basis of vector particle velocity (VPV) and acceleration (VPA).

For the purpose of analysis, geometry of selected mining panel has been modeled using NEi/NASTRAN computer code utilizing 3D FEM method. It was assumed that the overburden layers consists of several homogeneous rock plates reflecting the real lithology in the considered area. The averaged geological data and the estimated rock mass parameters are given in Table 1.

**Tab. 1. Geological data in the considered area**

| Rock type | Thickness (m) | $\sigma_c$ $\sigma_{cm}$ (MPa) | $\sigma_r$ $\sigma_{tm}$ (MPa) | $\sigma_t$ $\sigma_{sm}$ (Mpa) | $E_s$ $E^{(r)}$ (MPa) | Poisson's ratio $v$ | LEVEL |
|---|---|---|---|---|---|---|---|
| Basic Anhydrite | 73 | 95.5 / 19.53 | 5.47 / 0.24 | 15.28 / 3.12 | 54600 / 13650 | 0.25 | ROOF STRATA |
| Calcareous Dolomite III | 4 | 121.8 / 40.71 | 8.15 / 0.55 | 18.3 / 6.11 | 52900 / 13225 | 0.25 | |
| Calcareous Dolomite II | 10 | 149.2 / 63.96 | 9.60 / 0.89 | 22.38 / 7.64 | 58300 / 14575 | 0.26 | |
| Calcareous Dolomite I | 2 | 214.7 / 149.6 | 14.3 / 2.15 | 30.0 / 18.44 | 96200 / 24050 | 0.25 | |
| Copper ore | 2.4 | 116.4 | 6.9 | 19.8 | 16400 / 8200 | 0.23 | EXTRACTION RANGE |
| Grey Sandstone | 7 | 25.1 | 1.10 | 4.8 | 9400 / 4700 | 0.16 | FLOOR STRATA |
| Quartz Sandstone | 200 | 17.9 | 0.9 | 3.40 | 5100 / 2550 | 0.13 | |

- $\sigma_{cm}$, $\sigma_{tm}$, $\sigma_{sm}$ compression, tension and shear strengths for rock mass, assessed according to Hoek's [7] approach
- estimated laboratory results $\sigma_t = \omega \cdot \sigma_c$, where $\omega$ based on Lis [8] equals: 0.16 for anhydrites, 0.15 for dolomites, 0.19 for sandstones

The source of vibration was defined as simultaneous firing of explosives in 10 mining faces. Each face was treated as single vibration source. Detonation was simulated by applying of hydrostatic pressure into each wall of the solid finite elements increasing from 0 to 100 MPa at 1 ms and decreasing after detonation to zero after the next 1 ms.

Figure 2 shows the assumed function of the hydrostatic pressure (source of vibration) and method of simulation of explosive's detonation in numerical model.

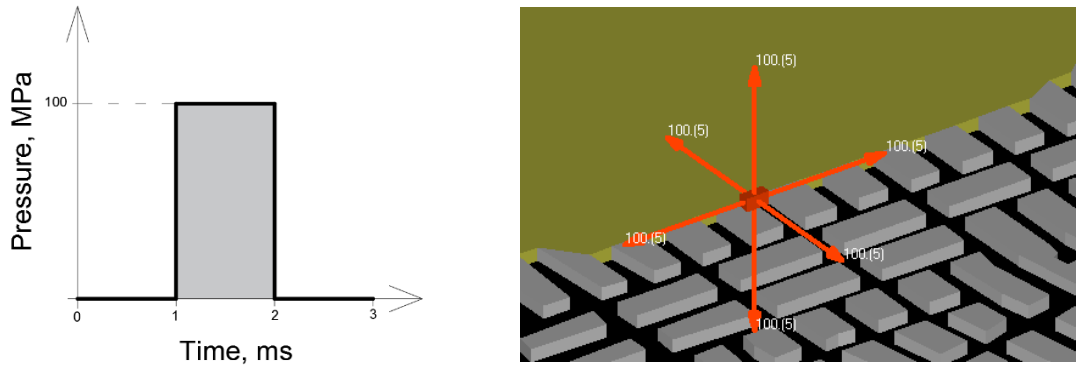

**Fig. 2. Assumed function of the hydrostatic pressure (left) and method of simulation of detonation (right) in numerical model**

Velocities and accelerations of paraseismic waves were calculates in three different points of the roof strata located in the forefield of the mining front, i.e. 40, 42 and 57 m from the face no. 5 (Fig. 3). All the points were situated in different roof levels, i.e. on the immediate roof strata (point #1), 14 m (point #2) and 40 m (point #3) above the roof surface. Geometry of the panel and location of analyzed points are presented in Figure 3.

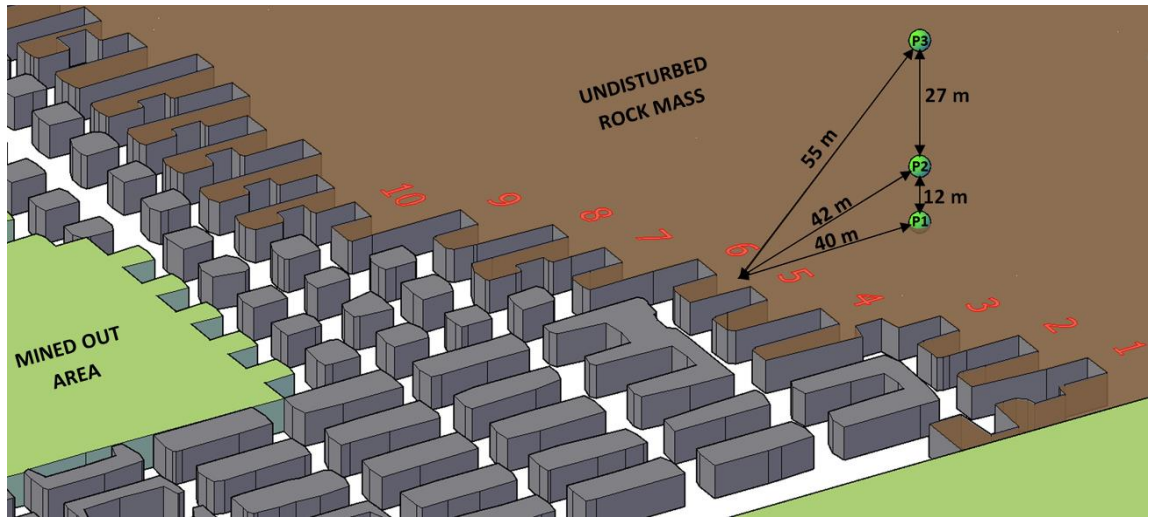

**Fig. 3. Geometry of mining panel used for numerical simulations**

### 3. ANALYSIS OF RESULTS

The analysis was done on the base of comparison of VPV in the time domain. Numerical calculation covers the period of time equal 0,25 s. (time step of 0.0005 s). For better visualization, results are shown in three separate charts for each specified points (Fig. 4).

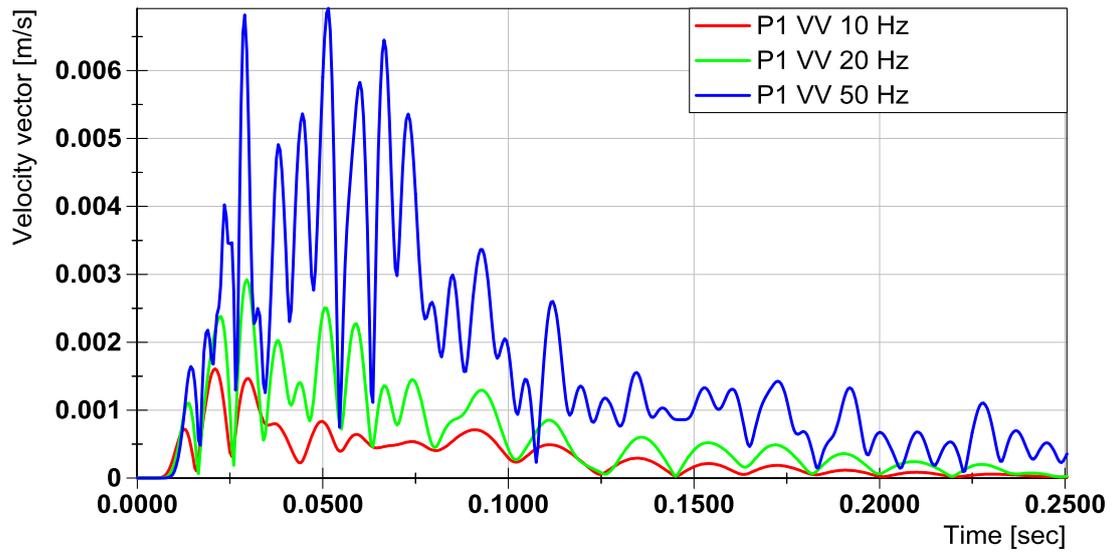
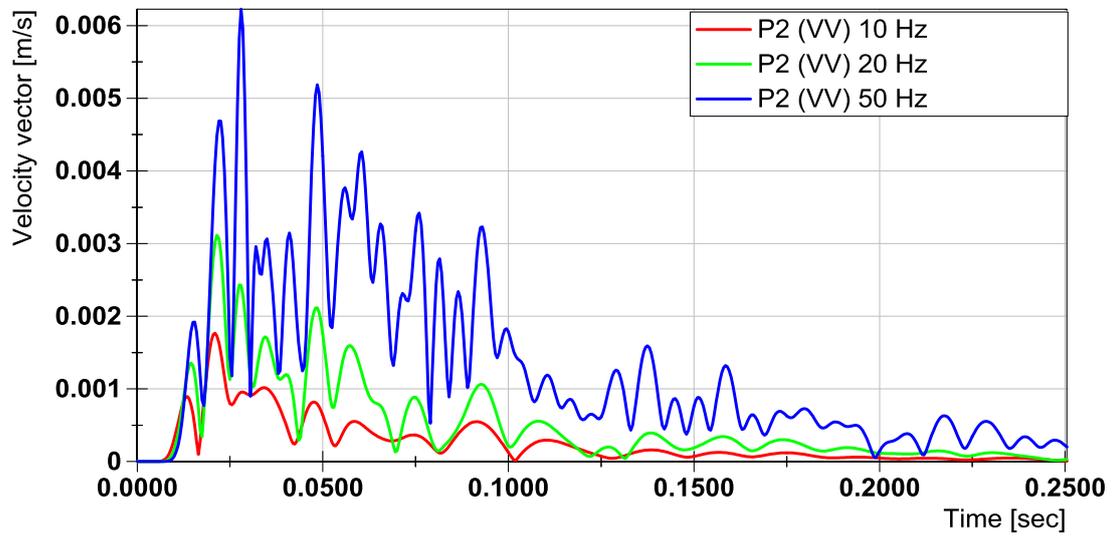
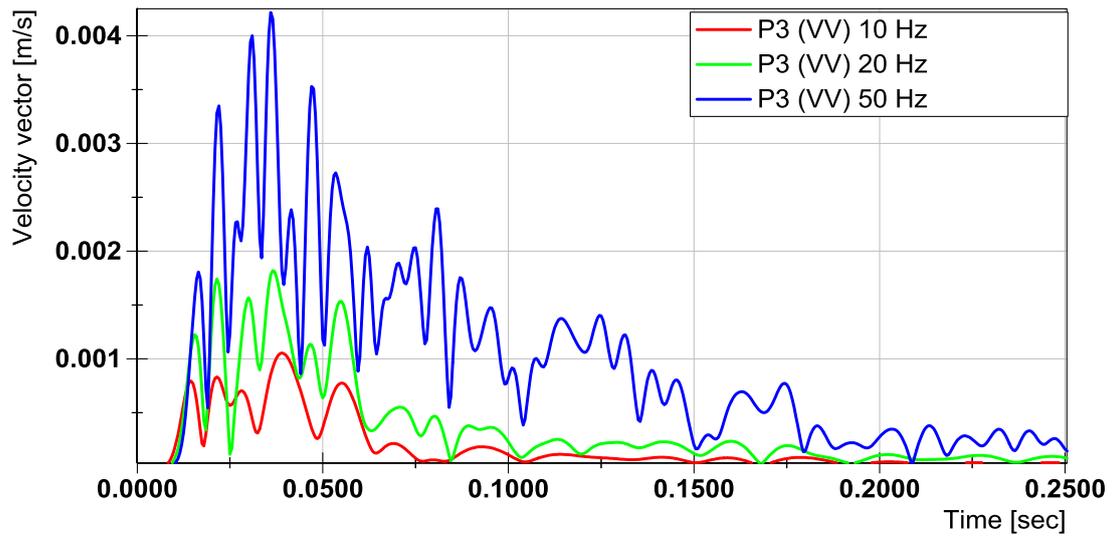

Fig. 4. Vector velocity vibrations for different frequencies and calculated points

Velocity vectors of vibrations for analysed time-dependent scenario show that the maximum value decreases when the distance is rising, as was expected. In the light of frequency, the simulations showed that maximum velocities are associated with 50 Hz for all selected points. Maximum value is about 7 mm/s for point #1 which is similar to the point #2 since the difference in the distances from the vibration sources measured in a straight line is relatively low for these two points. It should also be noted that there is significant difference between peak values for frequencies of 10 and 50 Hz what may suggest that simulation of the rock mass behaviour in the near-field strongly depends on the assumed frequency around which calculations of results were obtained. In addition to the analysis of the vibration velocity, acceleration in the same points within the roof strata were calculated. Figure 5 shows an example of acceleration in point #1 (immediate roof level) for all analysed frequencies.

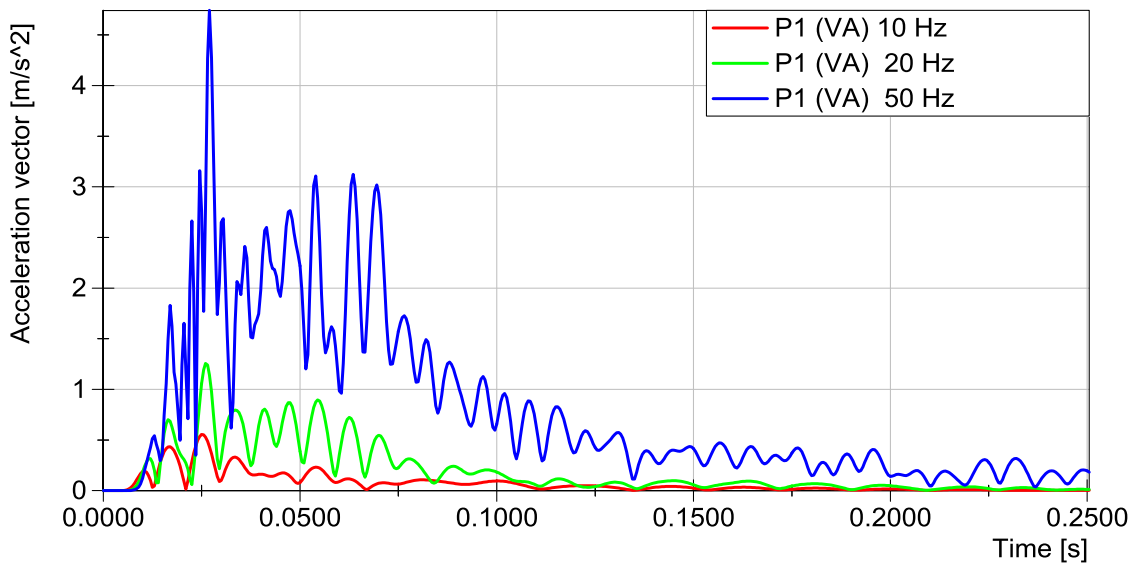

**Fig. 5. Vector acceleration vibrations for different frequencies and calculated points**

To summarize the simulations results, peak particle velocities for each frequencies and points are presented in Figure 6. Based on this one may conclude that the PPV value increases with the increase of assumed frequency value. However, changes of the PPV value for the frequencies above 50 Hz, should be analyzed.

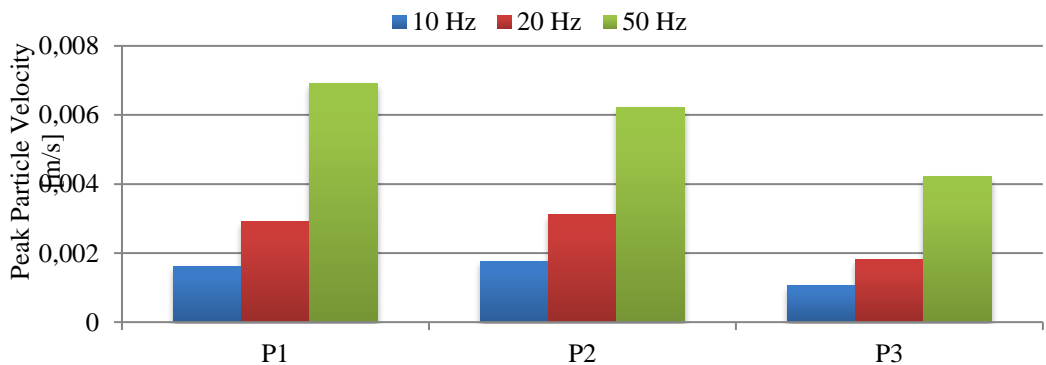

**Fig. 6. Peak particle velocity values for analyzed scenarios**

## 4. CONCLUSIONS

The effect of the frequencies' values assumed to be the dominant ones, around which the computer based 3D FEM solutions of the blasting operation has been simulated, reveals its significant importance in the peak particle values prediction at all of the selected points (nodes) located in the front of group of faces fired simultaneously. The time step value of Δt = 0.0005 ms applied in numerical analyses seems to be the adequate to obtain the time-dependent waveforms of ground velocities and acceleration, in the form of smooth and looking like correct lines.

On the other hand, the value of the overall structural damping coefficient assumed as G = 0.05 revealed to be too low to be able to suppress the vibration of the higher (50 Hz) frequency within rock mass at the distance of 40 ÷ 70 m. This indicates new fields in which more research work is necessary to be done.

The above presented work is the introduction to the problem of the optimal effect of group faces blasting, from production and safety points of view. Having significantly broader knowledge on the parameters governing the process, this activity would be directed towards the possibility of increasing of effectiveness the of blasting provocation by at least partial control of the frequency of induced seismic waves.

## ACKNOWLEDGEMENTS

**This paper has been prepared through the Horizon 2020 EU funded project on "Sustainable Intelligent Mining Systems (SIMS)", Grant Agreement No. 730302.**